
\magnification=1200\baselineskip=18pt
\bigskip \bigskip \centerline
{\bf Quantized Density Response in Insulators} \bigskip
\bigskip \bigskip \centerline
{\bf Qian Niu} \bigskip
\bigskip \bigskip \centerline
{\bf Department of Physics, University of Texas, Austin, Texas 78712} \bigskip
\bigskip \bigskip \centerline
{\bf Abstract} \bigskip

The response of particle density to a dilation of a periodic potential
in an insulator, with or without a fixed background potential or a magnetic
field,
is shown to be quantized.  A similar phenomenon occurs in a quantum Hall
system,
where the derivative of the electron density with respect to the magnetic field
is quantized in units of $e/h$, the inverse flux quantum.  A number of other
interesting results may be unified under the notion of quantized density
response,
including the quantization of adiabatic particle transport, the gap labeling
theorem
and its generalization in the presence of a magnetic field.
\vfil
\eject
\noindent
{\bf 1. Introduction}

It is well known that for a two dimensional gas of electrons
in a magnetic field $B$, the particle density $\rho$ satisfies $\rho/(eB/h)=n$
for $n$ filled Landau levels below the Fermi energy, where $h/e$ is the flux
quantum.
In the presence of a weak periodic potential, each Landau level
is split into $p$ subbands, if the magnetic flux in each
unit cell is $p\over q$ times the flux quantum, where $q$ and $p$ are
mutually prime integers.  When the Fermi energy lies in a gap between two such
subbands,
the ratio $\rho/(eB/h)$ is no longer an integer, because each filled subband
contains a
particle density $p$ times smaller than that for a full Landau level.
What remains to be quantized is instead the derivative
$${\partial\rho\over \partial(eB/h)}, \eqno (1)$$
which is also the integer for the quantized Hall conductance.[1-4]
Indeed, according to the Streda formula,[1] the Hall conductance may
be generally written as
$$ e\left(\partial \rho\over \partial B\right)_\mu,\eqno(2)$$
whenever the Fermi energy $\mu$ lies in an energy
gap (or even a mobility gap).   Under the same condition, the Hall conductance
and therefore
the density response are quantized.  Note that disorder is certainly allowed
under this
condition.

In this paper, we would  like  to show that the notion of quantized
density response may be applied to other type of insulators.
Consider now an electron gas in a periodic potential.  The average particle
density satisfies $\rho v=n$ for n filled Bloch bands, where $v$ is the unit
cell volume.  A natural question is whether one can generalize such a result to
the case with an additional background potential and/or a magnetic field.
The background potential may be periodic with a different  periodicity or even
be
disordered.  The magnetic field does not have to be uniform.
In analogy with the quantum Hall system discussed above, we would like to
expect
that the density response to a dilation of a periodic potential should be
quantized
when the Fermi energy lies in a gap (or even a mobility gap), although the
density
itself may not be quantized.  We show in this paper that this is indeed the
case for
a number of interesting situations.  We would like to emphasize that the
notion of quantized density response may provide a unifying framework in which
several important subjects, such as the quantum Hall effect, quantized
adiabatic
particle transport, and gap labeling theorem for quasi-periodic systems, are
found as
applications.

The organization of our paper is as follows.  Our initial study will be limited
to the relatively
simple case of one dimensional systems (Section 2), but the result will  be
generalized to two
dimensions afterwards (Section 3).  The presence of particle-particle
interactions makes the
problem complicated, but does not alter our conclusion at least for one
dimensional systems
(Section 4).  Density response to a magnetic field will then be studied in
Section 5.
Some useful applications of our results will be discussed in Section 6.

\bigskip
\noindent
{\bf 2. One Dimensional Case}
\bigskip
 Consider a
system of non-interacting electrons in the following potential at
zero temperature:
$$V(x)=V_1(\alpha x-\xi)+V_2(x),\eqno(3)$$
where $V_1$ is a periodic function of $\xi$ of period $1$, so that the
periodicity in $x$ is $1/\alpha$.  We will assume that the Fermi energy
lies in an energy gap for arbitrary $\xi$ and for a
small neighborhood of $\alpha$.  We would like to show that
${\partial \bar \rho\over\partial
\alpha}$ is quantized as integers, where $\bar \rho$ is the density
averaged over position in the bulk of
the system and over $\xi$ in a period (unity).  The average over $\xi$ will be
convenient for the mathematical manipulations, but is not really necessary if
the
potential $V_2$ is random, in which case there is no prefered position for
$V_1$.
The same is also true when $V_2$ is periodic but incommensurate with $V_1$.

\def\Tr #1 { {\rm Tr}\left\{ #1 \right \} }
The particle density at $x_0$ may be written as
$$\rho(x_0)=\oint {dz\over 2\pi i} \Tr {G \hat\rho(x_0)} , \eqno(4)$$
where $G=(z-H)^{-1}$ is the resolvent of the single particle
Hamiltonian, and $\hat \rho(x_0)=\delta(x-x_0)$ is the density operator.
The integral goes along a contour in the complex energy plane enclosing the
spectrum of filled states.  The contour will be chosen everywhere away from the
real axis
where the spectrum lies, except at the Fermi energy and a point below the whole
spectrum.
We assume that there are no extended states in the bulk at the Fermi energy.
It then follows that the Green function in the coordinate representation
$G(x,x')$ is
essentially local, i,e., it decays exponentially at large $|x-x'|$.[5]  Our
later manipulations
will depend critically on this property.

Using the
differential formula $dG=-GdG^{-1}G$, the density response can be written as
$${\partial\rho(x_0)\over\partial \alpha}=-\oint {dz\over 2\pi i} \Tr{
G{\partial
V_1(\alpha x-\xi)\over \partial\xi} x G \hat\rho(x_0)} .\eqno(5)$$
We have expressed the $\alpha$-derivative of $V_1$ in terms of its
$\xi$-derivative.
Next, we note that $xG$ may be written as
$G[G^{-1},x]G+G x=Gi\hbar  vG+G x$, where $v$ is the velocity
operator. Using further the fact that ${\partial V_1(\alpha x-\xi)\over
\partial\xi}
=-{\partial G^{-1}\over \partial\xi}$, we may write (5) as
$${\partial\rho(x_0)\over\partial \alpha}=-\oint {dz\over 2\pi i} \Tr{
G{\partial
V_1(\alpha x-\xi)\over \partial\xi}Gi\hbar v G \hat \rho(x_0) }
+{\partial \over \partial\xi} \oint {dz\over 2\pi i} \Tr{ G x \hat
\rho(x_0)} .\eqno(6)$$
When we average over $\xi$ in a unit cell, the second term on the
right hand side of (6) becomes zero because of the periodicity of $G$
in $\xi$.  In the following, we will no longer consider this term.

The locality of the Green function allows us to do a number of things.
First, the unboundedness of $x$ should not introduce any difficulty
in the above manipulations, if
we measure the density only in the bulk of the system.
Secondly, the locality of $G$ makes sure
that the expression (6) is insensitive to changes in the boundary
condition.  In particular, we may impose the quasi-periodic boundary
condition $\psi(x+L)=e^{i\theta}\psi(x)$, and make an average over $\theta$.
Under such a boundary
condition, the average of $\rho(x_0)$ over $x_0$ in the bulk may be
achieved by integrating over $x_0$ in the whole space and dividing it by $L$.
Therefore, with $\rho$ also averaged over $\xi$, we have
$${\partial\bar \rho\over\partial \alpha}
=-{1\over L}\int_0^{2\pi}{d\theta\over 2\pi}\int_0^1 d\xi\oint
{dz\over 2\pi i} \Tr{ G{\partial V_1(\alpha x-\xi)\over \partial\xi}Gi\hbar v
G} .\eqno(7)$$

To show explicitly the quantization of the density response,  we make a gauge
transformation
$\psi(x)\rightarrow\tilde\psi(x)= \psi(x) e^{-i\theta x/L}$, then
$G\rightarrow \tilde G$, and $             v\rightarrow -{L\over
\hbar}{\partial \tilde G^{-1}\over \partial \theta}$, and the
expression (7) may be written as
$$\eqalign{
&-\int_0^{2\pi} {d\theta\over 2\pi} \int d\xi \oint {dz\over 2\pi }
	\Tr {   \tilde G {\partial \tilde G^{-1}\over \partial\xi}
\tilde G {\partial \tilde G^{-1}\over \partial\theta} \tilde G  } \cr
=&\int_0^{2\pi}{d\theta\over 2\pi} \int d\xi \oint {dz\over 2\pi }
\Tr {    \tilde G{\partial  \tilde G^{-1}\over \partial\theta}
\tilde G{\partial  \tilde G^{-1}\over \partial\xi}     \tilde G } ,
\cr
}  \eqno(8)$$
where the second line is derived from the first by using the fact that
$\tilde G\tilde G=-{\partial \tilde G\over \partial z}$ and by integrating
over $z$ by parts.
According to Ref.[6], the above integral takes integer values and
represents a topological invariant for mappings from $T^3$ to nonsingular
operators.

Finally, we would like to remark that the condition that the Green function
falls off
exponentially at large distances is sufficient but not necessary.  A
sufficiently fast
power law fall off (e.g. $|x-x'|^{-3}$ in 1D) should still be alright for the
quantization of
density response in the thermaodynamic limit.  Such a remark also applies to
situations
to be discussed in the subsequent sections.

\noindent
{\bf 3. Two Dimensional Case}

The generalization of our result to higher dimensions involves
surprisingly a lot more algebra.  We have so far only achieved this for
two dimenions.  The
potential $V_1$ in (3) now has the form $V_1(\alpha_1x_1-\xi_1,
\alpha_2x_2-\xi_2)$, which is periodic in $\xi_1$ and $\xi_2$ with
unit periodicity.  We would like to show
that  $${\partial^2 \bar \rho\over\partial
\alpha_1\partial\alpha_2}\eqno(9)$$
is quantized as integers, where $\bar\rho$ is the density averaged over
the position in the bulk of the system and over $\xi_1$ and $\xi_2$.
Again, the average over the $\xi$'s is not really necessary if $V_2$ is random
or incommensurate with $V_1$.  We will also assume that there are no extended
states in the bulk at the Fermi energy, but states at the boundary, extended or
not,
are allowed. The system may also be in a magnetic field, which may or may not
be uniform.  The locality of the Green function will again be the key to our
problem.

The general method is quite parallel to the one dimensional case.
Equation (5) now becomes
\def\r{{\bf r}}
$$\eqalign{
{\partial^2\rho(\r_0)\over\partial \alpha_1\partial
\alpha_2}=\oint {dz\over 2\pi i} {\rm Tr}\bigl\{ \hat\rho(\r_0)
&[G{\partial V_1\over \partial\xi_1} x_1 G{\partial V_1\over \partial\xi_2}
x_2G\cr
+&G{\partial V_1\over \partial\xi_2} x_2 G{\partial V_1\over \partial\xi_1}
x_1G\cr
+&G{\partial^2 V_1\over \partial\xi_1\partial\xi_2} x_1 x_2 G] \bigr\}
 .\cr}\eqno(10)$$
The term including the double derivative of $V_1$ may be rewritten as
$$-G{\partial V_1\over \partial\xi_1} G x_1x_2{\partial V_1\over \partial\xi_2}
G -G x_1x_2{\partial V_1\over \partial\xi_2}G{\partial V_1\over
\partial\xi_1}G,\eqno(11)$$
where we have ignored a term of total derivative in $\xi_1$, because such a
term will
become zero when we average over $\xi_1$ in (9).  The quantity in the square
bracket in
(10) may now be written as
$$\eqalign{
&G{\partial V_1\over \partial\xi_1} [x_1, G]{\partial V_1\over \partial\xi_2}
x_2G
+Gx_2{\partial V_1\over \partial\xi_2} [G, x_1]{\partial V_1\over
\partial\xi_1}G\cr
=&G{\partial V_1\over \partial\xi_1}Gi\hbar v_1G{\partial V_1\over
\partial\xi_2} x_2G
-Gx_2{\partial V_1\over \partial\xi_2} Gi\hbar v_1G{\partial V_1\over
\partial\xi_1}G,\cr}\eqno(12)$$
where we have
used the fact that  $[x_1,G]=G[G^{-1},x_1]G=Gi\hbar v_1G$. Next, we may replace
$x_2 G$ in
the first term  and $Gx_2$ in the second term of the last expression by
$\pm[x_2,G]=\pm
Gi\hbar v_2G$ respectively, because the error made in (10) is proportional
to a total derivative in $\xi_1$ and $\xi_2$:
$$\eqalign{
{\rm Tr} \bigl\{\hat \rho(\r_0)[
&   G {\partial V_1\over \partial\xi_1} Gv_1G {\partial V_1\over \partial\xi_2}
Gx_2
-x_2G {\partial V_1\over \partial\xi_2} Gv_2G {\partial V_1\over \partial\xi_1}
G ]\bigr\} \cr
={\rm Tr}\bigl \{\hat \rho(\r_0) x_2  [
&{\partial \over\partial \xi_1} (Gv_1{\partial G\over \partial\xi_2})
-{\partial \over\partial \xi_2} (Gv_1{\partial G\over \partial\xi_1})  ]\bigr\}
,\cr
}\eqno(13)$$
where we have used the cyclic property of the trace and the commutativity of
$x_2$
and $\hat\rho(\r_0)$.  As a result of these considerations, we may now rewrite
the quantity
in (10) averaged over $\xi_1$ and $\xi_2$ as  $$ \int\int d\xi_1 d\xi_2\oint
{dz\over 2\pi i}
\Tr{ \hat\rho(\r_0) [G{\partial V_1\over \partial\xi_1}Gi\hbar v_1 G{\partial
V_1\over
\partial\xi_2}Gi\hbar v_2G +Gi\hbar
v_2G{\partial V_1\over \partial\xi_2}Gi\hbar v_1 G{\partial V_1\over
\partial\xi_1}G]}
\eqno(14)$$
Finally, as in the one dimensional case, we introduce the quasiperiodic
boundary conditions in
both $x$ and $y$ directions, and make average over the phase angles $\theta_1$
and
$\theta_2$, then the quantity in (9) becomes
$$\eqalign{ \int\!\!\int d\xi_1 d\xi_2
\int\!\!\int {d\theta_1d\theta_2\over (2\pi)^2}
\oint {idz\over 2\pi }
{\rm Tr}\bigl \{
 & \tilde G {\partial \tilde G^{-1}\over \partial\xi_1} \tilde G
            {\partial \tilde G^{-1}\over \partial\theta_1}
   \tilde G {\partial \tilde G^{-1}\over \partial\xi_2} \tilde G
            {\partial \tilde G^{-1}\over \partial\theta_2} \tilde G  \cr
+& \tilde G {\partial \tilde G^{-1}\over \partial\theta_2} \tilde G
            {\partial \tilde G^{-1}\over \partial\xi_2}
   \tilde G {\partial \tilde G^{-1}\over \partial\theta_1}\tilde G
            {\partial \tilde G^{-1}\over \partial\xi_1}\tilde G
  \bigr\}  ,
\cr} \eqno(15)$$
where we have made a gauge transformation so that the phase angle dependence in
the boundary
conditions goes explicitly into the Hamiltonian.

In the appendix we show that (15) can be further written as
$$\eqalign{
\int\!\!\!\int\!\!\!\int\!\!\!\int\!\!\!\oint{ id\xi_1
d\xi_2d\theta_1d\theta_2dz\over
480\pi^3} {\rm Tr }
\bigl\{ &\tilde G {\partial \tilde G^{-1}\over \partial\xi_1}
		\tilde G {\partial \tilde G^{-1}\over \partial\theta_1}
		\tilde G {\partial \tilde G^{-1}\over \partial\xi_2}
		\tilde G {\partial \tilde G^{-1}\over \partial\theta_2}
		\tilde G {\partial \tilde G^{-1}\over \partial z}\cr
+&{\rm 119\  antisymmetrizing\ terms}\bigr\}.\cr}
\eqno(16)$$
According to Ref.[7], such an expression takes integer values and represents
the winding
number for the mapping from $T^5$ to nonsingular operators $G$.  Therefore, the
density
response (9) is quantized as integers.

\bigskip
\noindent {\bf 4. The Case With Many Body Interactions}
\bigskip

In this section we try to generalize the result of quantization of density
response
to the case with many body interactions.  This can be readily done by invoking
the
relationship with quantized particle transport (see the next section), which
has been
established in the presence of many-body interactions.  As we show below, such
a
generalization can also be done directly within the present formalism in one
dimension.

Eq.(5) remains valid if we regard $G$ as the Green function for the many body
Hamiltonian,
regard $\hat\rho(x_0)$ as $\sum_{j=1}^N \delta(x_j-x_0)$, and replace
${\partial V_1(\alpha
x-\xi)\over \partial \xi} x$  by $\sum_{j=1}^N {\partial V_1(\alpha
x_j-\xi)\over \partial \xi} x_j$, where $j$ labels the particles.  The contour
integral is now
surrounding the ground state  energy of the many body system.
Eq.(6) will also remain valid if we make similar changes
on the right hand side; explicitly, the second term becomes
$$ \oint {dz\over 2\pi i}\sum_j\Tr{ G{\partial V_1(\alpha
x_j-\xi)\over \partial \xi}G x_j \hat\rho(x_0)} , \eqno(17)$$
where the trace is over the antisymmetric many-body states only.
In the coordinate representation, $G(x_1,x_2,\cdots,x_N;
x_1',x_2',\cdots,x_N')$, the Green
function is antisymetric under permutation of each set of position variables.
We may thus
replace $x_j$ in the above equation by $\sum x_k/N$ under the trace operation.
 Then Eq.(17)
becomes a total derivative in $\xi$, and can therefore be ignored.
Similarly, Eq.(7) now becomes $${\partial\bar \rho(x_0)\over\partial \alpha}
=-{N\over L}\int_0^{2\pi}{d\theta\over 2\pi}\int d\xi\oint
{dz\over 2\pi i} \sum_j\Tr{ G{\partial V_1(\alpha x_j-\xi)\over
\partial\xi}Gi\hbar v_j
G} .\eqno(18) $$
Again, because of the antisymmetry of the Green functions in particle labels,
we may
replace $v_j$ by $\sum_k v_k/N$.  The final result of (8) can then be
rederived,
showing the quantization of density response in the presence of many body
interactions.

Unfortunately, similar arguments in the two dimensional case show that the
density response
is given by the expression (15) or (16) divided by $N$, if we regard $G$ as the
many-body
Green function.  This says that density response is only quantized in integer
units of $1/N$.

\bigskip
\noindent
{\bf 5. Density Response to a Magnetic Field}
\bigskip
\def \r {{\bf r}}
\def \p {\bf p} \def \A {\bf A}
Having studied the density response to the dilation of periodic potentials, we
now consider
the problem of density response to a change of a magnetic field.  Associated
with the magnetic
field is a length, $\sqrt{\hbar\over eB}$, the so called magnetic length, whose
square gives
the area occupied by one flux quantum.  Our problem can therefore also be
stated as how the
electron density is controled by the scale of the magnetic length rather than
by the unit
cell length scale of a periodic potential.  As we pointed out before, this is a
problem closely
related to the quantum Hall effect, but our treatment will be shaped in a way
more parallel to the
method used in the previous sections than those used by other authors.[8]

Consider a two dimensional electron gas in a uniform magnetic
field $B$ in the  perpendicular direction.   The single electron Hamiltonian
may be written as
$$ H={1\over 2m} (\p+e\A)^2+V(\r)  \eqno (19) $$
where $\A$ is the vector potential and may be taken as $(A_1, Bx+A_2)$ with
$A_1$ and $A_2$
being constants independent of $B$.  We have no special requirement
on the scalor potential $V(\r)$.  The particle density is still given by a
formula like in (4)
except that $\rho(x_0)$ should be replaced by the corresponding  2D quantity
$\rho(\r_0)$.  Since
the magnetic field dependence comes only from the Green function,  the
differential density
response to the magnetic field can be calculated using the  formula
$dG=-GdG^{-1}G$ as
$${\partial\rho(\r_0)\over\partial B}=-\oint {dz\over 2\pi i} \Tr{
G {\partial H\over \partial A_2} x G \hat\rho(\r_0)} \eqno(20)$$
Next, we note as before that $xG$ may be written as
$G[G^{-1},x]G+G x={i\hbar\over e} G{\partial H\over \partial A_1} G+G x$.
Equation (20) can then be written as
$${\partial\rho(\r_0)\over\partial B}=-{\hbar\over e} \oint {dz\over 2\pi }
\Tr{
G {\partial H\over \partial A_2} G{\partial H\over \partial A_1} G
\hat\rho(\r_0)} \eqno(21)$$
where we have dropped a term proportional to the $z$-integral of $\Tr{
G {\partial H\over \partial A_2} G x\hat\rho(\r_0) } $.  Such a term is
proportional to $x_0$
times  the derivative of $\rho(\r_0)$ with respect to $A_2$, and is zero
because of gauge
invariance.

As before, the z-integral goes along a contour in the complex energy plane
enclosing the
spectrum of filled states.  The contour will be chosen everywhere away from the
real axis
(where the spectrum lies), except at the Fermi energy and a point below the
whole spectrum.
We assume that there are no extended states in the bulk at the Fermi energy.
It then follows that the Green function in the coordinate representation
$G(\r,\r')$ is
essentially local, i,e., it decays exponentially at large $|\r-\r'|$.[5]  We
can thus
impose periodic boundary conditions to the wave functions, making only
exponentially small errors
in the density response in the bulk.  We can then average over $r_0$ on the
boundaryless torus,
yielding
$${\partial\rho(\r_0)\over\partial B}=-{\hbar\over eL_1L_2} \oint {dz\over 2\pi
} \Tr{
G {\partial G^{-1}\over \partial A_2} G{\partial G^{-1}\over \partial A_1} G }
\eqno(22)$$
Also upto exponentially small errors, we can make a further average over $A_1$
and $A_2$ in the
intervals $(0,{h\over eL_1})$ and $(0,{h\over eL_2})$ corresponding to unit
flux quanta through
the holes of the torus, with the result
$${\partial\rho(\r_0)\over\partial B}=-{e\over h} \oint {dz\over (2\pi)^2
}\int\int dA_1 dA_2 \Tr{
G{\partial G^{-1}\over \partial A_2} G{\partial G^{-1}\over \partial A_1} G }
\eqno(23)$$
which is essentially the same as Eq.(1) of Ref.[6], because  the above
expression is antisymmetric
upon interchanging $A_1$ and $A_2$, and because the $z$-derivative of $G^{-1}$
is unity.  The
density response to the magnetic field is therefore quantized in units of
$e/h$.

As has been shown in the context of the quantum Hall effect, the above result
allows a number of
generalizations.  First, the system does not have to be strictly two
dimensional.  Also,
one can add a modulation to the magnetic field without changing the
quantization of the
density response to the uniform part of the field.  Further, one can
reformulate the
above derivations using many-body quantities (cf. Section 4), allowing
interactions between
electrons.  Finally, when the many-body state does not go back to itself when a
flux quantum is
inserted through a hole of the torus, which occurs in a fractional quantum Hall
regime, one
obtains  fractionally quantized density response to the field.

\noindent
{\bf 6. Applications}

In one dimension, quantized density response to a dilation of a periodic
potential is the same as
the quantized adiabatic particle transport studied by Thouless and others.[8,5]
 When $\alpha$ is
increased by $\delta\alpha$, the  potential $V_1$ shrinks a distance of
$\delta\alpha L \over
2\alpha$ at $x=\pm L/2$.  If the particle transport is $t$ when the potential
is shifted by a
period  ($1/\alpha$), there will be $t\alpha\delta\alpha
L/(2\alpha)=t\delta\alpha L/2$ particles
transported inward through the point $x=L/2$, and there will be the same amount
of particles
transported inward through  the point at $x=-L/2$.  The density increase is
therefore
$t\delta\alpha$. In fact, the expression (7) is identical to that for the
particle transport.
However, the concept of  density response can be readily generalized to more
than one dimension.

The quantized density response as a topological invariant has an important
physical content: the system may be classified into topologically distinct
situations, two
situations being topologically the same if one can be continuously transformed
into the other
without closing the gap at the Fermi energy.  Two situations should
have the same integer of density response if they are topologically identical.
This follows
from the fact that the expression for the density response is a continuous
functional of the
Hamiltonian, and from the fact that the value of the density response is
discrete.

One special class is for $V_1=const.$ or weak, where we have to assume that
$V_2$ is such
that there is always a gap at the Fermi energy. A $V_1$ potential is considered
weak as long
as it does not close the energy gap at the Fermi energy as it is turned on from
zero.
The integer of density response for this class is zero.

Another special class is for $V_2=const.$ or weak.  For this class we only need
to
consider the case of $V_2=0$. The density response for $n$ filled bands is
simply $n$,
because there is one electron per unit cell for each filled band, discounting
the spin
degeneracy.

More interesting situations occur when the potential is a superposition of
periodic ones.
Consider the following example in one dimension:
$$V(x)=\sum_j V_j(\alpha_jx-\xi_j), \eqno(24) $$
where each $V_j$ has period 1 in $\xi_j$.  We assume that the gap at the Fermi
energy does
not close for a neighborhood of $\alpha_j$, and that $t_j$ is the integer of
density
response to the $j$th potential, then
$$\delta \bar\rho=\sum_j t_j \delta \alpha_j.\eqno (25)$$
Under a uniform scaling of $x\to (1+\delta\alpha)x$, we have $\delta \bar
\rho=\bar\rho \delta\alpha$.  On the other hand, such a scaling is equivalent
to setting
$\delta \alpha_j=\alpha_j \delta \alpha$, so that $\delta \bar \rho= \delta
\alpha \sum_j t_j\alpha_j$, according to (25).  Comparison of the two results
yields
$$\bar \rho=\sum_j t_j\alpha_j. \eqno (26)$$
A similar result has been obtained by Thouless using quantized particle
transport.[9]
In higher dimensions, the present theory allows us to generalize the above
result to
$$\bar \rho=\sum_j t_j/v_j,\eqno (27)$$
where $v_j$ is the unit cell volume of the $j$th potential.[10]

In two dimensions and in the presence of a magnetic field $B$, one has to add
to the right
hand side of (27) an extra term $\sigma e\delta B/h$, where $\sigma$ is the
integer for the
quantum Hall effect.  Under a uniform scaling of $\delta \alpha$, we have to
also set $\delta
B=\delta\alpha B$, then we have
$$\bar \rho=\sum_j t_j/v_j+\sigma eB/h. \eqno (28) $$
Such a result was first derived by Wannier in the study of the Harper's
equation[11], and has been proved in general by Dana et al using magnetic
translation
group.[4]  A three dimensional generalization of the above relation has been
studied
in Ref.[12].

There are two remaining problems to be solved.  The first is to generalize the
result of
density response quantization to three (and higher) dimensions.  It may be
possible to
do so within the present formalism, but the algebra involved will be quite
heavy.  The
second is to show (or disprove) the quantization in two or more dimensions with
particle
particle interactions present.  It is possible that fractionally quantized
density response
may occur for the interacting case.[13]

\noindent {\bf Acknowledgment}

The author wishes to thank X. G. Wen,
D. J. Thouless, Y. Avron and E. Akkermans for useful discussions, and to thank
the hospitality of
the Institute for Theoretical Physics at Technion, Israel, where part of  this
work was done.
This work was supported by the Welch Foundation and by an NIST Precision
Measurement Grant.
\vfil
\eject

 \bigskip \noindent
\centerline {\bf Appendix}
\bigskip

To simplify notations, we represent the quantity in the square brackets in (15)
symbolically as
$$\xi_1\theta_1\xi_2\theta_2+\theta_2\xi_2\theta_1\xi_1. \eqno(A1)$$
The original problem is symmetric under the exchange of $(1\leftrightarrow 2)$,
(A1) is
equivalent to half of
$$\eqalign{
&\xi_1\theta_1\xi_2\theta_2+\theta_2\xi_2\theta_1\xi_1\cr
 +&\xi_2\theta_2\xi_1\theta_1+\theta_1\xi_1\theta_2\xi_2.\cr} \eqno(A2) $$
By integrating $z$ by parts, we can rewrite (A1) as
$$\eqalign
{&-\theta_1\xi_2\theta_2\xi_1-\xi_2\theta_2\xi_1\theta_1-\theta_2
\xi_1\theta_1\xi_2\cr
&-\xi_2\theta_1\xi_1\theta_2-\theta_1\xi_1\theta_2\xi_2-
\xi_1\theta_2\xi_2\theta_1. \cr}
\eqno(A3)$$
Therefore, (A2) is equal to
$$\eqalign
{&-\theta_1\xi_2\theta_2\xi_1-\theta_2\xi_1\theta_1\xi_2\cr
 &-\xi_2\theta_1\xi_1\theta_2-\xi_1\theta_2\xi_2\theta_1. \cr}
\eqno(A4)$$
It then follows that (A1) may be written as $1\over 4$ of (A2) plus (A4), i.e.
$$\eqalign{
 &\xi_1\theta_1\xi_2\theta_2-\xi_1\theta_2\xi_2\theta_1\cr
+&\theta_2\xi_2\theta_1\xi_1-\theta_2\xi_1\theta_1\xi_2\cr
+&\xi_2\theta_2\xi_1\theta_1-\xi_2\theta_1\xi_1\theta_2\cr
+&\theta_1\xi_1\theta_2\xi_2-\theta_1\xi_2\theta_2\xi_1.\cr} \eqno(A5) $$

The first line of (A5) may be written as
$$\eqalign{
&\tilde G {\partial \tilde G^{-1}\over \partial \xi_1}
{\partial \tilde G\over \partial \theta_1}
{\partial \tilde G^{-1}\over \partial \xi_1}
{\partial \tilde G\over \partial \theta_2}
-\tilde G {\partial \tilde G^{-1}\over \partial \xi_1}
{\partial \tilde G\over \partial \theta_2}
{\partial \tilde G^{-1}\over \partial \xi_1}
{\partial \tilde G\over \partial \theta_1} \cr
=-&{\partial \tilde G\over \partial \theta_1}
		{\partial \tilde G^{-1}\over \partial \xi_1}
	\tilde G
{\partial \tilde G^{-1}\over \partial \xi_1}
{\partial \tilde G\over \partial \theta_2}
+ {\partial \tilde G\over \partial \theta_2}
{\partial \tilde G^{-1}\over \partial \xi_1}
\tilde G
{\partial \tilde G^{-1}\over \partial \xi_1}
{\partial \tilde G\over \partial \theta_1}, \cr }\eqno(A6)$$
where we have integrated $\theta_1$ by parts in the first term,
and $\theta_2$ in the second term. Written in abstract notation, (A6) is
$\theta_2\xi_1\xi_2\theta_1-\theta_1\xi_1\xi_2\theta_2$.
Similar manipulations can be made on the remaining lines of (A5),
showing that (A5) is equal to
$$\eqalign {
 &\theta_2\xi_1\xi_2\theta_1-\theta_1\xi_1\xi_2\theta_2 \cr
+&\theta_1\xi_2\xi_1\theta_2-\theta_2\xi_2\xi_1\theta_1 \cr
+&\xi_2\theta_1\theta_2\xi_1-\xi_1\theta_1\theta_2\xi_2 \cr
+&\xi_1\theta_2\theta_1\xi_2-\xi_2\theta_2\theta_1\xi_1. \cr }\eqno(A7)$$

Next, by integrating $z$ by parts, we can show that (A7) is further equal
to
$$\eqalign {
 &\xi_1\xi_2\theta_2\theta_1-\xi_1\xi_2\theta_1\theta_2 \cr
+&\xi_2\xi_1\theta_1\theta_2-\xi_2\xi_1\theta_2\theta_1 \cr
+&\theta_1\theta_2\xi_2\xi_1-\theta_1\theta_2\xi_1\xi_2 \cr
+&\theta_2\theta_1\xi_1\xi_2-\theta_2\theta_1\xi_2\xi_1. \cr }\eqno(A8)$$

In summary, we may write (A1) as $1\over12$ of (A5)+(A7)+(A8), namely
$$\tilde G {\partial \tilde G^{-1}\over \partial\xi_1} \tilde G
{\partial \tilde G^{-1}\over \partial\theta_1}
\tilde G {\partial \tilde G^{-1}\over \partial\xi_2}\tilde G
{\partial \tilde G^{-1}\over \partial\theta_2}\tilde G+23{\rm\ antisymetrizing\
terms,}$$
where `antisymmetrizing' acts on the four variables $\xi_1,\theta_1,\xi_2,$ and
$\theta_2$.
Then, we append each term a factor of ${\partial \tilde G^{-1}\over \partial
z}=1$, and
use the cyclic property of the trace, yielding the expression (16) in the text.

\vfil
\eject

\bigskip  \noindent {\bf REFERENCES}  \bigskip
\item {1.} P. Streda, {\it J. Phys. C} {\bf 15}, L717  (1982).
\item {2.} P. Streda, {\it J. Phys. C} {\bf 15}, L1299 (1982).
\item {3.} D. J. Thouless, M. Kohmoto, M. P. Nightingale, and M. den Nijs,
{\it Phys. Rev. Lett.} {\bf 49}, 405 (1982).
\item {4.} I. Dana, Y. Avron, and J. Zak, {\it J. Phys. C} {\bf 18}, L679
(1985);   J. E. Avron and L. G. Yaffe, {\it Phys. Rev. Lett.} {\bf 56}, 2084
(1986).
\item {5.} Q. Niu and D. J. Thouless, {\it J. Phys. A} {\bf 17}, 2453 (1984).
\item {6.} H. Aoki and T. Ando, {\it Phys. Rev. Lett.} {\bf 57}, 3093 (1986).
\item {7.} E. Witten, {\it Nucl. Phys. B} {\bf 223}, 422 (1983);  R. Bott and
R. Seeley, {\it Commun. Math. Phys.} {\bf 62}, 235 (1978).  The base manifolds
considered in these references are spheres instead of tori, but the
quantization of an expression like Eq.(16) is valid independent of the
topology of the base manifold.
\item {8.} J. E. Avron, R. Seiler, and B. Simon, {\it Phys. Rev. Lett.} {\bf
65}, 2185 (1990). J. Bellisard, In proceedings of the Bad-Schandau Conference
on
Localization, edited by W. Weller and P. Zieshe (Teubner, Leipzig, 1988); J.
Xia, Comm. Math. Phys. 119, 29-50 (1988).
\item {9.} D. J. Thouless, {\it Phys. Rev. B} {\bf 27}, 6083 (1983); Q. Niu,
Phys. Rev. B34 (1986) 5093. Also see Ref.[5].
\item {10.} R. Johnson and J. Moser, {\it Commun. Math. Phys.} {\bf 84}, 403
(1982).
\item {11.} G. H. Wannier, {\it Phys. Status Solidi} {\bf 88}, 757 (1978)
\item {12.} G. Montambaux and M. Kohmoto, {\it Phys. Rev. B} {\bf 41}, 11417
(1990); M. Kohmoto , B. I. Halperin, and Y-S. Wu, {\it Phys. Rev. B} {\bf 45},
13488 (1992). \item {13.} See Section 5 in the context of quantum Hall effect,
and see Q. Niu, Mod. Phys. Lett. B5, (1991) 923, in the context of adiabatic
particle transport. \vfil
\eject
\end